\documentclass[letter]{aa}     

\usepackage{graphicx}
\usepackage{txfonts}
\usepackage{lipsum}
\usepackage{subcaption}         
\usepackage{lscape}             
\usepackage{placeins}           
                                
\usepackage{natbib}
\usepackage{sidecap}
\usepackage{orcidlink}
                                
\usepackage[dvipsnames]{xcolor}

\usepackage{hyperref}
\hypersetup{
	colorlinks=true,
	linkcolor=Cerulean,
	citecolor=Cerulean
}

\begin{document}

   \title{On the origin of the rotation of massive stars}

   \author{André Oliva\inst{1,2} \orcidlink{0000-0003-0124-1861}
   	\and Facundo D. Moyano\inst{3,1} \orcidlink{0000-0002-9633-4093}
   	\and Luca Sciarini\inst{1} \orcidlink{0009-0001-7331-2238}
   	\and Sylvia Ekström\inst{1}\orcidlink{0000-0002-2564-5660}
   	\and Patrick Eggenberger\inst{1}\orcidlink{0000-0001-6319-9297}
   	\and Georges Meynet\inst{1}\orcidlink{0000-0001-6181-1323}
   	 \fnmsep
        }

   \institute{Département d'Astronomie, Université de Genève, Chemin Pegasi 51, 1290 Versoix, Switzerland\\
             \email{andre.oliva@ucr.ac.cr}
            \and Space Research Center (CINESPA) and School of Physics, University of Costa Rica, Ciudad Universitaria Rodrigo Facio, 11501 San José, Costa Rica
            \and Yunnan Observatories, Chinese Academy of Sciences, Kunming 650216, China}

   \date{Received September 30, 20XX}

  \abstract  
   {We explore the origin of the rotation rates of massive stars. Contrary to their low-mass siblings, most massive stars do not have detectable magnetic fields, so that star-disk interaction models used for the formation of rotating low-mass stars do not apply.}
   {We investigate whether the magnetic fields of protostellar jets present in the parent molecular cloud prevent the protostar from reaching the critical angular velocity.} 
   {Starting from the gravitational collapse of a molecular cloud, we run two two-dimensional radiation-gravito-magnetohydroynamical simulations to study the formation of an accretion disk and the launching of magnetically-driven protostellar outflows (of particular interest is the formation of a magnetocentrifugal jet originating from the protostar and inner disk). We then study the angular momentum transfer from the disk and jet onto the protostar. Finally, we compute one-dimensional stellar evolution models of the pre-main sequence including our results from the disk-jet simulations and follow the angular momentum redistribution within the structure of the protostar.}
   {We find that the angular momentum transported outwards by the magnetically-driven protostellar outflows is sufficient for keeping the protostar below the critical speed at all times. Moreover, we are able to link the strength of the jet, and thus the rotation rate at the end of the accretion epoch, to the initial conditions for star formation. Our results show that the jet strength produces a variety of stellar rotation rates, suggesting that protostellar jets fix the rotation rate of massive stars.}
   {}


   \maketitle
   \nolinenumbers


\section{Introduction}
The origin of the rotation of massive stars is still unclear.
Surveys of rotation rates of O and B type stars typically show a bimodal distribution of rotation rates \citep{Holgado2022, Britavskiy2023, Huang2008, Huang2010}, where fast rotators are proposed to be formed by binary interaction \citep{Mink2013}.
\cite{Holgado2022} and \cite{Huang2010} have estimated that the initial distribution of rotation rates of O and B type stars at the beginning of the main sequence to be around 10\% to 20\% of the critical rotation rate (i.e., the rotation rate at which  gravito-centrifugal equilibrium is achieved).
However, this critical rotation rate is reached very quickly when modeling protostars accreting through a disk \citep{2017A&A...602A..17H} and during protostellar binary mergers \citep{2020A&A...644A..41O} and thus angular momentum removal mechanisms must exist during the pre-main-sequence evolution.

Modeling the origin and evolution of stellar spins requires a coordinated description of the (proto)stellar interior, its surrounding material, and how angular momentum is transferred from the latter to the former.
\cite{2005ApJ...632L.135M, 2008ApJ...678.1109M, 2008ApJ...681..391M} proposed accretion-powered winds as a solution to the stellar angular momentum problem in the case of low-mass stars.
More recently, \cite{2019A&A...632A...6G, 2023A&A...678A...7A} and \cite{2023A&A...673A..54G} linked star-disk interaction simulations by \cite{2022ApJ...929...65I} to stellar evolution models, although they did not include a detailed treatment of angular momentum transport within the star (instead, they assumed a solid-body instantaneous redistribution).
Moreover, those models require the presence of large-scale stellar magnetic fields.
Only about 10\% of massive stars exhibit such magnetic fields during their main-sequence evolution \citep[see, e.g.][]{2009ARA&A..47..333D}.
Furthermore, an external convective envelope able to sustain a solar-like dynamo is not necessarily present during the whole pre-main-sequence evolution of a massive star due to strong changes in the stellar structure \citep[see][]{1996A&A...307..829B, 2009ApJ...691..823H, 2017A&A...602A..17H}.
Therefore, the principal mechanism responsible for angular momentum removal from the protostar while in the embedded phase must work independently of any coupling of the stellar magnetic field (if it exists) and the accretion disk.
This is consistent with the findings of \cite{Holgado2022}, where no strong magnetic braking effects are observed in massive stars throughout their main-sequence evolution, that is, once the natal cloud is dispersed.

In the context of low-mass star formation, simulations of star-disk interaction including disk- and stellar winds such as \cite{2005ApJ...635L.165R, 2013A&A...550A..99Z, 2015ApJ...798..116R, 2022ApJ...929...65I} typically start from an initial model for the accretion disk and protostellar outflows.
In recent years, however, it has been possible to model the formation of the disk and protostellar outflows starting directly from the gravitational collapse of a molecular cloud \citep{2023A&A...669A..81O, 2021A&A...656A..85M} including comparison against observations \citep{2024A&A...690A..81M, 2025A&A...697A.206S}, although an accurate description of the innermost $\sim 3\,\mathrm{au}$ in the cloud remains challenging until now.

In this letter, we study the evolution of the spin of massive (proto)stars by performing and connecting two different computer models: a disk-jet \emph{ab-initio} simulation (Section \ref{s: amjet}) and a stellar evolution simulation (Section \ref{s: stellar am}). 
To this end, we first calculate the mass and angular momentum transfer exerted by the disk and jet onto the protostar, and then we introduce our results in the stellar evolution model.
Our main goal is to test whether magnetocentrifugally driven protostellar jets are able to keep the protostar rotating well below the critical velocity, without any additional assumptions on the presence of large-scale stellar magnetic fields or of other stellar winds.
A discussion of our results and conclusions are offered in Section \ref{s: conclusions}.

\section{Angular momentum removal by a protostellar jet} \label{s: amjet}

\begin{figure*}
	\centering
	\includegraphics[width=0.9\textwidth]{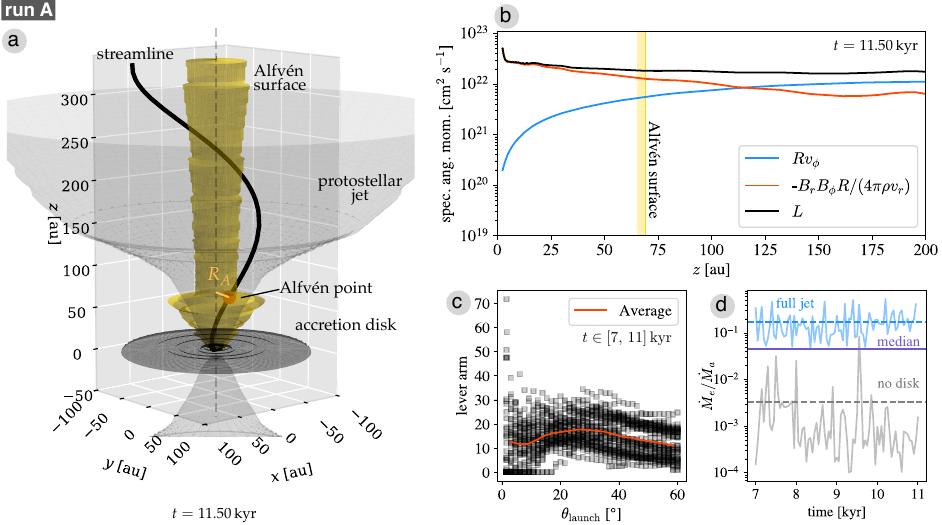}
	\caption{Overview of the angular momentum transport done by the streamlines of the magneto-centrifugal jet, using the data from run A. (\emph{a}) Three-dimensional view of one streamline launched from the inner disk at $R_{lp}=3\,\mathrm{au}$, traced using the velocity field that co-rotates with the launching point. (\emph{b}) Specific angular momentum transported by the streamline depicted in Panel \emph{a}. (\emph{c}) Lever arm $\Lambda$ of jet streamlines launched at $r=3\,\mathrm{au}$ from different launching angles within a time window. (\emph{d}) Ratio $\xi$ of the mass ejected through the Alfvén surface to the accreted mass. The calculation of the ejected mass for the light-blue line includes all streamlines of the jet and the gray line excludes those originated at the surface of the disk. The thick purple line indicates the median value considering the results of both calculations.} \label{f: streamline}
\end{figure*}

We performed two simulations of the formation of the accretion disk and the magnetically-driven jet starting from the collapse of a slowly-rotating molecular cloud (which we call the disk-jet \emph{ab-initio} simulations).
The circumstellar material is modeled using magnetohydrodynamics, including diffusive radiation transport, Ohmic dissipation and self-gravity.

The simulations start from identical initial conditions, such that we obtain an accretion rate of $\approx 5\cdot 10^{-4}\,\mathrm{M_\odot\,yr^{-1}}$. 
We previously used one of the simulations (which we call run A) to trace detailed streamlines of the magnetocentrifugal jet originating in the inner disk, leading to strong agreement with observations of water masers in the star-forming region IRAS 21078+5211 \citep{2022NatAs...6.1068M}.
Run A has an inner boundary of radius $3\,\mathrm{au}$, and we assume that matter that enters this region is accreted by the protostar.
In order to check the convergence of our results and understand the mass and angular momentum transfer from the inner disk onto the surface of the star, we reduced the inner boundary in run B down to a radius of $r = 150\mathrm{\,R_\odot} $ $ \approx 0.7\mathrm{\,au}$ \footnote{Hereafter, $r$ is the spherical radial coordinate and $R$ is the cylindrical radial coordinate.}.
This new reduced value is within the order of magnitude expected for the radius of the bloated phase of the protostar (see e.g. \citealt{1996A&A...307..829B}, \citealt{2009ApJ...691..823H}).
More details on the setup of the simulations can be found in Appendix \ref{s:disk-jet simulations}, and a convergence study is offered in Appendix \ref{s: convergence}	.

\subsection{Mass and angular momentum flow along streamlines} \label{ss: streamline}

We compute the angular momentum carried away by the jet by following several of its streamlines within a time window.
To illustrate the process, we first discuss how one of such streamlines removes angular momentum from one point in the inner disk, and then we repeat the process for several launching points surrounding the protostar.

We start by focusing on the streamline shown in Fig. \ref{f: streamline}a, which was computed in the frame of reference co-rotating with the launching point, that is, the vector field $\vec v - \omega_\text{lp} R \vec e_\phi$ ($\vec v$ being the velocity, $\omega_\text{lp}$ the angular velocity of the launching point, $R$ the cylindrical radius and $\vec e_\phi$ the unit vector in the azimuthal direction).
The material is launched from the surface of the inner disk at a distance of $R_\text{lp}=3\mathrm{\,au}$ from the protostar and at a time $t=11.50\mathrm{\,kyr}$ after the start of the simulation.\footnote{We selected this particular example because the jet is free of transient features and the streamline is launched from the smallest possible spherical radius that our computational domain allows to study directly.}
In this region, the magnetic field is strong enough and the density low enough for the flow to become sub-Alfvénic (this corresponds to the interior of the yellow surface in Fig. \ref{f: streamline}a; see a more detailed discussion in \citealt{2023A&A...680A.107M, 2024A&A...690A..81M} and \citealt{2023A&A...669A..81O}).
As a consequence, the ejecta is forced to follow a magnetic field line while inside of the Alfvén surface and when viewed from the co-rotating frame of reference.

Along a streamline, the quantity
\begin{equation} \label{e: L along streamline}
	L = r v_\phi - \frac{B_r}{4\pi \rho v_r}r B_\phi
\end{equation}
is a constant of motion if the magnetic field only has radial and azimuthal components (see Appendix \ref{s: windtorque} for a discussion on how to arrive at this expression).
We have checked in the simulations that in the jet, $B_r$ dominates over $B_\theta$ and $B_\phi$ (see \citealt{2023A&A...669A..81O}).
To check that this quantity is a constant of motion also in our simulations, we calculate $L$ and present the results in Fig. \ref{f: streamline}b.
The specific angular momentum $rv_\phi$ increases as the ejecta are accelerated along the field line while the magnetic torque decreases, keeping $L$ approximately constant along the streamline.

When the flow accelerates and becomes super-Alfvénic, it becomes capable of dragging the magnetic field lines with it, and thus angular momentum can be transferred onto large scales originating from the launching point.
The torque $\mathrm dJ/\mathrm dt$ done by the magnetic outflow is then computed by (see e.g. \citealt{Maeder2009} \S 13.2)
\begin{equation}\label{e: wind torque}
	\frac{\mathrm dJ}{\mathrm dt} = \frac{2}{3}\frac{\mathrm dM}{\mathrm dt}R_\text{lp}^2 \Omega_\text{lp} \left( \frac{R_A}{R_\text{lp}} \right)^2,
\end{equation}
where $\mathrm dM/\mathrm dt \equiv \dot M_e$ is the mass flow that crosses the Alfvén surface at a cylindrical radius $R_A$ (see Fig. \ref{f: streamline}a), $R_\text{lp}$ is the cylindrical radius of the launching point and $\Omega_\text{lp}$, its angular velocity. 
The ratio $R_A/R_\text{lp} \equiv \Lambda$ is known as the magnetic lever arm and expresses the amplification of the angular momentum outflow due to the magnetic torque.
For the streamline shown in Fig. \ref{f: streamline}, we obtain that the Alfvén point is located at $R_A = 26.8\mathrm{\,au}$, yielding a lever arm of $8.93$.

We repeat this process for many launch points situated at a distance of $3\,\mathrm{au}$ away from the protostar (limited by the inner boundary of our run A), using several launch angles, within the time window $t\in [7,11]\,\mathrm{kyr}$, and compute the lever arm, the results of which are shown in Fig. \ref{f: streamline}c.
The time average of the lever arm is between 10 and 20 for the different launch angles.
In producing this result, however, we made the following simplification: since we do not need the azimuthal position of the Alfvén point, we traced only the poloidal magnetic field lines and computed their intersection with the Alfvén surface.

In our model of the torque, we write the ejected mass flow as a fraction $\xi$ of the accretion rate $\dot M_a$, that is, $\dot M_e \equiv \xi \dot M_a$.
We integrated the mass flux that exits through the Alfvén surface ($\int \rho \vec v \cdot d\vec A $, where $d\vec A$ points outwards with respect to the Alfvén surface) for the same time window, resulting in an average mass outflow rate of $\approx$17\% the mass inflow rate passing through the disk (see Fig. \ref{f: streamline}d).
This calculation sums all the mass lifted by the jet, including the mass launched from the disk surface and the surroundings of the protostar.
We also repeated the calculation by excluding mass from streamlines originating directly on the disk surface, yielding a time average of 0.3\%.
This value has to be considered a lower bound for $\xi$ because our inner boundary condition which prevents mass from leaving the central sink cell where the protostar forms.
The median value taking into account the results of both calculations is $\xi = 4.5$\%.
We take this value as a reference for our models.

\subsection{Discussion: extension of our calculations to the stellar surface} \label{ss: discussion one}

Strictly speaking, the surface of the protostar is not included within the computational domain of the disk-jet simulations.
We pose the question, how applicable are these ideas to the angular momentum transfer onto and from the surface of the forming protostar?

In low-mass stars, the torque resulting from the interaction of the (poloidal) stellar magnetic field and the magnetic field of the disk must be considered.
Because in massive protostars the conditions for producing a stable solar-like dynamo are not fulfilled during their whole pre-main-sequence evolution, we simply do not include in our models any poloidal stellar magnetic field and examine instead the question of whether the magnetic fields of the jet and the cloud alone are already able to exert a sufficiently strong torque to keep the protostar below the critical surface rotation rate at all times.
The scenario we investigate is similar to the accretion-powered wind described in \cite{2005ApJ...632L.135M}, but we propose that the cloud magnetic field determines the Alfvén radius $R_A$ instead of the stellar magnetic field, and thus the lever arm can be ultimately linked to the conditions of the natal environment of the star (i.e., the properties of molecular clouds).
This is because the gravitational collapse of the cloud produces naturally the kind of strong radial magnetic fields around the protostar that are at the core of how \eqref{e: wind torque} was derived, in a similar way that a dipolar stellar magnetic field does it in a magnetized stellar wind.

\section{Angular momentum evolution of the protostar} \label{s: stellar am}

\begin{figure}
	\centering
	\includegraphics[width=0.9\columnwidth]{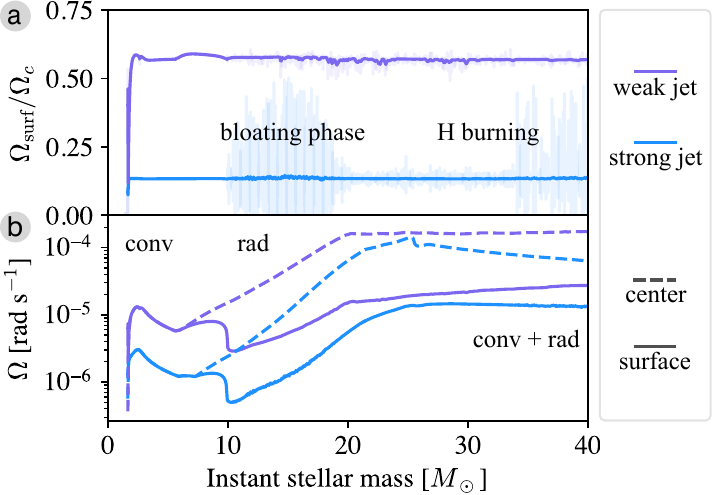}
	\caption{(\emph{a}) Evolution of the rotation criticality as a function of the instantaneous stellar mass under two different jet parametrizations. (\emph{b}) Evolution of the surface and central angular velocity of the protostar as the structure changes from fully convective to fully radiative and then a convective core and radiative envelope. The opaque lines are running averages.} \label{f: surf am}
\end{figure}

The angular momentum accretion and its redistribution along the structure of forming massive star was investigated using the Geneva stellar evolution code ({\tt Genec}, see \citealt{2008Ap&SS.316...43E}).
We modified the code to include the angular momentum extraction done by the jet according to \eqref{e: wind torque} as well as the angular momentum delivered by the Keplerian-like accretion disk.
We started the calculations from the pre-main-sequence evolution of a protostar, and sustained the accretion of mass and angular momentum even after the onset of hydrogen burning, as expected during the formation of massive stars (see e.g. \citealt{2017A&A...602A..17H, 2010ApJ...722.1556K}).

Motivated by the results of our \textit{ab-initio} disk-jet simulations, we set a constant accretion rate of $\dot M_a = 5\cdot 10^{-4} \,\mathrm{M_\odot/yr}$ and chose two parametrizations of the jet, both with a lever arm of $\Lambda=20$: a ''strong'' jet of ejection-to-accretion rate $\xi=0.05$ and a ''weak'' jet with $\xi=0.01$ (cf. Fig. \ref{f: streamline}).
Other values consistent with our disk-jet simulations will be discussed in future work; we have chosen only those two cases in order to explain the main idea.
Once inside of the star, the accreted angular momentum is transported diffusively by convection, horizontal turbulence and the secular shear instability.

The evolution of the surface angular velocity of the (proto)star is presented in Fig. \ref{f: surf am}.
We define the rotation criticality as $\Omega_\text{surf}/\Omega_c$. The critical angular velocity is computed as in \cite{Ekstroem2012} (taking into account the deformation of the star as it spins), but for context, it is roughly $\Omega_c = \sqrt{GM_\star/R_\star^3}$.
Our results show that rotation criticality is determined by the strength of the jet, and remains constant in time for a given jet strength.
This does not mean that the surface angular velocity is constant in time  (Fig. \ref{f: surf am}b), because $\Omega_c$ also changes in time.

The rotation criticality is found to be independent of the stellar radius and the stellar structure.
Even during the strong changes in stellar radius occurring during the bloating phase, the rotation criticality does not change.
We interpret this result to be a consequence of the disk and jet torques reaching a rotational equilibrium very quickly, and the fact that we have used constant values of $\xi$ and $\Lambda$.
In Appendix \ref{s: genecmodels}, we discuss this situation more carefully using a simple analytical argument, and present further details on the setup of the stellar evolution simulations.

Our models reveal that even though a spin-up and a spin-down happen during changes in the moment of inertia of the star (for example during the bloating phase), the surface angular momentum is dominated by the accretion of the incoming material.
The difference between the surface and stellar angular velocity (Fig. \ref{f: surf am}b) shows how the accreted angular momentum is then redistributed very efficiently in the case of a fully convective structure  (resulting in solid body rotation) and less efficiently with the presence of a radiative zone (resulting in differential rotation).

\section{Discussion and conclusions} \label{s: conclusions}

The idea that an important part of the angular momentum removal is done by accretion-powered winds is not new (\citealt{2019A&A...632A...6G, 2023A&A...678A...7A, 2008ApJ...678.1109M, 2008ApJ...681..391M, 2023A&A...673A..54G}, etc.).
However, previous work focused on low-mass star formation, and thus poloidal stellar magnetic fields were considered, which determine the magnetic lever arm (see e.g. \citealt{2022ApJ...929...65I}).
From our results, we argue that the magnetic field of the cloud (i.e. the magnetic field of the jet) has by itself an adequate structure and capacity for producing sufficient magnetic braking to keep the protostar below the critical rotation rate at all times as long as there is accretion (or that the same magnetic field structure remains attached to the stellar surface).
Moreover, Fig. \ref{f: surf am} shows that in principle, with a sufficiently strong jet, one can obtain any surface rotation rate at the end of the star formation process.
The jet strength is a direct result of the initial conditions of the gravitational collapse of the cloud \citep{2023A&A...669A..81O}, so we can link the rotation rate of a massive star just after the formation process directly to its natal environment.
According to our results, during the accretion epoch and just after it finishes, the rotation rate of massive stars is in principle not correlated to stellar properties such as mass.

However, in reaching this conclusion we do not consider the possible additional effects of spin-up or spin-down due to binary interactions, nor the additional spin-down effects expected from the interaction of any existing (proto)stellar magnetic field and the disk.
Additionally, our setup is two-dimensional, so we neglect three-dimensional effects in angular momentum transport (breaking of axial symmetry due to disk fragmentation or turbulence).

In future work, we will address the dependence of the angular momentum removal done by the jet with the initial conditions for star formation, how this determines the rotation rate of the formed massive star and what is the role of stellar feedback (which may lead to dependencies of our results with metallicity).

\begin{acknowledgements}
	AO has received support from ESKAS No 2023.0405 and the University of Costa Rica project No 829-C6-133.
	LS and SE have received support from the SNF project No 212143. PE have received support from the SNF project No 219745.
\end{acknowledgements}

\bibliographystyle{aa} 
\bibliography{bibliography}

\begin{appendix}

\section{Disk-jet simulations} \label{s:disk-jet simulations}

\begin{figure}
	\centering
	\includegraphics[width=\columnwidth]{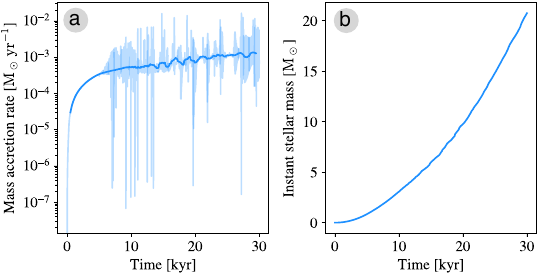}
	\caption{(\emph{a}) Mass accretion rate and (\emph{b}) stellar mass as a function of time using the data from run A.}
	\label{f: accrate}
\end{figure}

The disk-jet simulations start from the collapse of a gravitational cloud initially containing $100\,\mathrm{M_\odot}$ enclosed within a radius of $0.1\,\mathrm{pc}$.
This initial configuration produces an accretion rate of $\approx 5 \cdot 10^{-4}\,\mathrm{M_\odot\,yr^{-1}}$ including the effects of the outflows (see Fig. \ref{f: accrate} for more details).
The motion of the gas is modeled with the methods of non-ideal magnetohydrodynamics using the {\tt Pluto} code \citep{2007ApJS..170..228M} with the module {\tt Haumea} for self-gravity \citep{2010ApJ...722.1556K} and the \cite{2007ApJ...670.1198M} prescription for Ohmic resistivity as a function of density only (with a fixed temperature of $10\mathrm{\,K}$).
Radiation transport is taken into account using the gray flux-limited diffusion approximation using the module {\tt Makemake} as described in \cite{2020ApJS..250...13K}. More details on the methods used in the disk-jet simulations can be found in \cite{2023A&A...669A..80O} and \cite{2023A&A...669A..81O}.

Initially, the density in the cloud is distributed according to the power law $\rho \propto r^{-1.5}$. It is slowly rotating as a solid body with its total rotational energy being 4 \% of its total gravitational energy.
There is an initially-uniform magnetic field of $68.31\,\mu\mathrm{G}$ oriented parallel to the rotation axis, which corresponds to a mass-to-flux ratio of 20 times the critical (collapse-preventing) value.

We used a time-independent, two-dimensional, axisymmetric grid in spherical coordinates: the radial coordinate increases logarithmically with distance from the center, and the polar coordinate increases linearly from the rotation axis until the midplane, where we assume equatorial symmetry.
Two different grids were used in this study.
The first case (run A) corresponds to the fiducial run with the grid labelled as x8 in \cite{2023A&A...669A..81O}\footnote{It also corresponds to Simulation 1 in \cite{2023A&A...680A.107M}}, that is, a radial coordinate that extends from $r=3\,\mathrm{au}$ to $r=0.1\,\mathrm{pc}$ divided in 448 cells, and a polar angle divided in 80 cells.
In the second case (run B), the grid extends from $r=0.698\,\mathrm{au}\approx 150\,\mathrm{R_\odot}$ to $2500\,\mathrm{au}$, while maintaining the same initial density distribution than in the first case.
In run B, there are 160 cells in the radial coordinate and 32 cells in the polar coordinate.

The goal of this study is to understand the angular momentum removal done by the magnetically-driven outflows, so no stellar feedback is taken into account.
This approximation is valid for the early stages of the formation of a massive star.
We will address the effects of stellar feedback in a future study, but we anticipate that they will increase the angular momentum removal done by the total protostellar outflow and thus help further reducing the rotation rate of the (proto)star.
At any point in time, the disk-jet simulation computes the instantaneous stellar mass as the mass that has entered the sink cell, and interpolates the stellar evolution tracks of \cite{2009ApJ...691..823H} to compute the stellar radius.
This is the way we obtained the stellar radii quoted in Sect. \ref{s: amjet}, but not in Sect. \ref{s: stellar am}, which were computed directly with the Geneva stellar evolution code.

\section{Convergence of our results with sink cell size} \label{s: convergence}

\begin{figure}
	\includegraphics[width=\columnwidth]{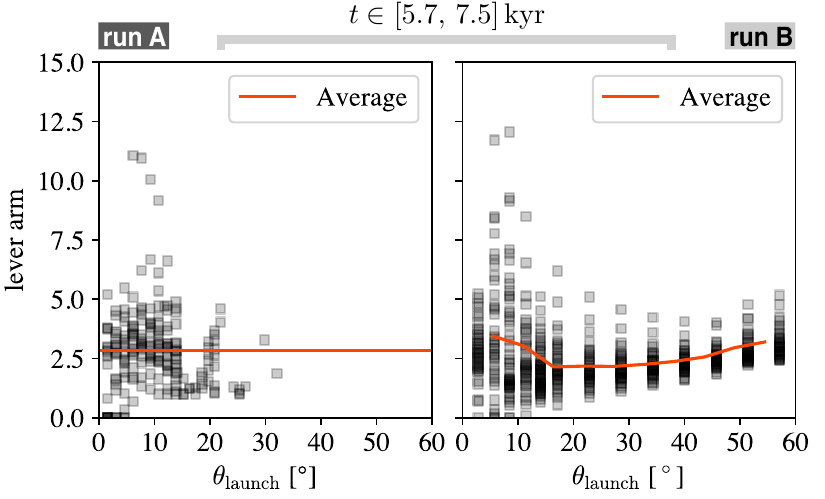}
	\caption{Convergence of the lever arm with sink cell size} \label{f: convergence}
\end{figure}

Since the inner boundary of run A is located at $r=3\,\mathrm{au}$, far away still from the stellar surface, we produced run B in order to study the convergence of the results of Sect. \ref{ss: streamline} when the launching points get closer to the star.
A reduced sink cell size dramatically increases computing time due to the decrease in the orbital timescale in the inner disk.
This meant that we were able to compute the evolution of the disk-jet system only up to the initial stages of jet launch, that is, until 7.5 kyr after the onset of the gravitational collapse.
At this stage, the Alfvén surface has not yet grown to its full size, and thus the Alfvén radii are smaller than in the time window considered in the rest of the work.

We repeated the calculations of the lever arm for several launch angles around the protostar, for both run A and run B, in the time window $t\in [5.7,\, 7.5]\,\mathrm{kyr}$ (the results are shown in Fig. \ref{f: convergence}).
The reduction in sink cell size made for run B does not significantly affect the average value of the lever arm.
In run A, the jet is narrower than in run B (therefore, there are only launching points for $\theta_\text{launch} \lesssim 30^\circ$ in run A).
In run B in turn, the smaller sink cell means that the inner disk is better resolved in that time window, and thus the jet is wider in comparison to run A as the disk also launches streamlines into the jet.
We conclude from this exercise that the lever arm values deduced from run A with a launch radius of $3\,\mathrm{au}$ hold for smaller launch radii, and therefore it is reasonable to use them to compute the wind torque down to the stellar surface.

\section{Angular momentum transported by a magnetized wind} \label{s: windtorque}

Consider an axisymmetric, fully ionized plasma with density $\rho$, velocity $\vec v = v_r \vec e_r + v_\phi \vec e_\phi$ and magnetic field $\vec B = B_r \vec e_r + B_\phi \vec e_\phi$ surrounding a star (subindex $\star$).
In the steady state, the momentum equation in the azimuthal direction is
\begin{equation}
	\rho [(\vec v \cdot \nabla) \vec v]_\phi = \frac{1}{4\pi}[(\nabla \times \vec B) \times \vec B]_\phi
\end{equation}
\begin{equation} \label{e: mphi}
	\implies \frac{d}{dr}(rv_\phi) = \frac{B_r}{4\pi \rho v_r} \frac{d}{dr}(rB_\phi)
\end{equation}
The solenoidality of the magnetic field ($\nabla \cdot \vec B = 0$) implies that
\begin{equation} \label{e: solB}
	\frac{1}{r^2}\frac{\partial}{\partial r} (r^2 B_r) = 0 \implies r^2 B_r = \text{const},
\end{equation}
and the conservation of mass implies that
\begin{equation} \label{e: consmass}
	r^2 \rho v_r = \text{const}.
\end{equation}
One can use \eqref{e: solB} and \eqref{e: consmass} to integrate \eqref{e: mphi} to obtain a constant of motion
\begin{equation}
	rv_\phi - \frac{B_r}{4\pi \rho v_r} r B_\phi = \text{const} \equiv L.
\end{equation}
The first term in this equation is the specific angular momentum and the second term is the torque exerted by magnetic stresses.
Using the frame which co-rotates with the launching point, and introducing the definition of the Alfvén speed, one can find $L$ to be \citep{1967ApJ...148..217W}
\begin{equation}
	L = \Omega_\text{surf} R_A^2,
\end{equation}
where $\Omega_\text{surf}$ is the angular velocity of the surface of the star (which is the same angular velocity as the launching point) and $R_A$ is the radius where the flow switches from sub-Alfvénic to super-Alfvénic.
From this, we conclude that the torque done onto the star by the flow should have the form
\begin{equation} \label{e: simpletorque}
	\frac{dJ_\star}{dt} \sim \Omega_\text{surf} R_A^2 \frac{dM}{dt}
\end{equation}
as proposed in, for example, \cite{1967ApJ...148..217W}, \citet[\S 13.2]{Maeder2009}.
The full derivation of this expression (see \citealt{Mestel1984}) contains an integral of the streamlines originating in the surface of the star and which cross the Alfvén surface (to account for the different latitudes of origin, we multiply an extra geometrical factor of $2/3$ like in \citealt{1967ApJ...148..217W,2005ApJ...632L.135M}).
Even though some streamlines of the jet originate close to the poles, they can still carry a significant angular momentum flow, as long as the Alfvén radius reached is sufficiently large.

\section{Stellar evolution models} \label{s: genecmodels}

\begin{figure}
	\centering
	\includegraphics[width=0.7\columnwidth]{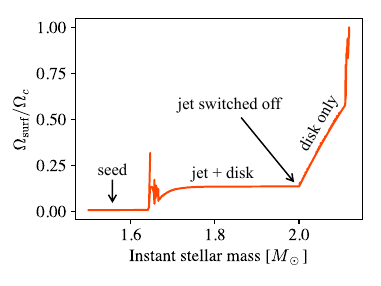}
	\caption{Example of the stellar angular momentum problem using {\tt Genec}: after switching off the jet, the protostar reaches very quickly the critical speed.}
	\label{f: amprob}
\end{figure}

\begin{figure}
	\centering
	\includegraphics[width=\columnwidth]{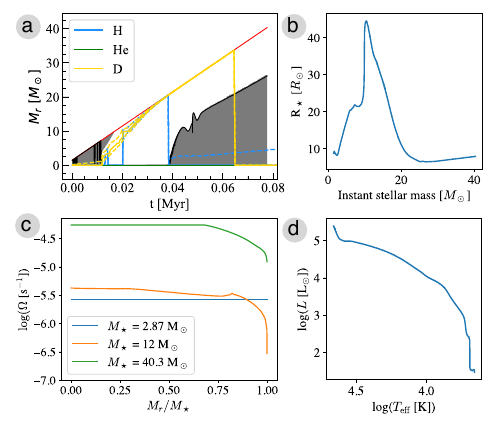}
	\caption{Overview of the evolution of the protostar. (\emph{a}) Kippenhahn diagram showing deuterium and hydrogen burning, with the red line being the total instantaneous stellar mass and the shadowed regions indicate convection. An element is considered to be burning when the energy generation rate is at least $10^2\,\mathrm{erg\,g\,s^{-1}}$. The dotted lines indicate the limits of the region where the corresponding element is burned, and the solid lines indicate the center of the burning region for each element (colors). (\emph{b}) Evolution of the stellar radius. (\emph{c}) Internal distribution of the stellar angular velocity for different times (expressed as instant stellar masses). (\emph{d}) Hertzsprung--Russell diagram of the accreting protostar (the bottom right corresponds to the initial time).} \label{f: evolution}
\end{figure}

We modified the Geneva stellar evolution code to include the (spin-down) torque done by the protostellar jets using \eqref{e: wind torque}. Introducing the definitions of $\xi$ and $\Lambda$, and taking the launching point as an average across the surface of the star (see \citealt{2019A&A...632A...6G}), we obtain the prescription
\begin{equation} \label{e: Jjet}
	\dot J_\text{jet} = \frac{2}{3} \xi \dot M_a  \Lambda^2 R_\star^2 \Omega_\text{surf}, 
\end{equation}
where we take $R_\star = \sqrt{\mathcal L_\star/(4\pi\sigma)}/T_\text{eff}^2$ ($\mathcal L_\star$ being the stellar luminosity, $\sigma$ the Stefan-Boltzmann constant and $T_\text{eff}$ the effective temperature) and $\Omega_\text{surf}$ is the angular velocity of the surface of the star just before it is entrained by the jet \footnote{As a check, we have computed the torque of the jet as it leaves the Alfvén surface in the disk-jet simulation (as described in Sect. \ref{ss: streamline}, and discarding the streamlines that originate in the disk) and the torque obtained with the semi-analytical formula \eqref{e: Jjet}. At $t = 11.50\mathrm{\,kyr}$ (same considered in Fig. \ref{f: streamline}) both formulas yield a torque of the order of $10^{42}\,\mathrm{g\, cm^2\, s^{-2}}$.}.
The (spin-up) torque delivered by the disk onto the protostar is included as \citep{2017A&A...602A..17H}
\begin{equation} \label{e: acctorque}
	\dot J_\text{disk} = \dot M_a\sqrt{GM_\star R_\star}.
\end{equation}
The net torque onto the stellar surface is then $\dot J = \dot J_\text{disk} - \dot J_\text{jet}$.
As noted by \cite{2005ApJ...632L.135M}, a consequence of writing the ejection rate as proportional to the accretion rate, and the Alfvén radius as proportional to the stellar radius is the following.

If we assume that the mass and radius of the star change slowly, then the accretion torque \eqref{e: acctorque} is roughly constant.
However, the jet torque \eqref{e: Jjet} depends on $\Omega_\text{surf}$, which means that there is a value of $\Omega_\text{surf}$ for which $\dot J_\text{disk} = \dot J_\text{jet}$, that is, rotational equilibrium is produced.
The equality of the torques leads very easily to the expression
\begin{equation}
	\Omega_\text{surf}= \frac{3}{2}\frac{1}{\xi \Lambda} \Omega_c,
\end{equation}
which means that a constant $\xi$ and $\Lambda$ produce a constant rotation criticality.
In our disk-jet and stellar evolution models, we wish to test whether we reach a rotational equilibrium where $\Omega_\text{surf} < \Omega_c$.

Rotation inside of the star is treated with the shellular rotation hypothesis, that is, that the angular velocity is constant on isobars.
The angular momentum is transported within the star with a diffusive scheme, neglecting advection by meridional currents.
We include the diffusion of angular momentum by horizontal turbulence with the diffusion coefficient of \cite{Zahn1992} (which depends on the velocity of the meridional current).
Additionally, we consider the effects of secular shear turbulence with the diffusion coefficient by \cite{Maeder1997}.
A large diffusion coefficient models the efficient angular momentum transport in convective regions \citep{2008Ap&SS.316...43E}.

We start the calculation with the adiabatic contraction of a protostellar seed of $1.5\,\mathrm{M_\odot}$ in very slow rotation forming from material with solar metallicity (i.e., star formation in the present-day universe).
After some tests we will present in future work, we concluded that the mass and angular velocity of the seed do not significantly change our results; this is in agreement with previous findings by \cite{2023A&A...673A..54G}.

As a test of the stellar angular momentum problem, we ran the model with a weak jet and switched it off at an arbitrary time, in this case just after the protostar has reached a mass of $2\,\mathrm{M}_\odot$.
The results of this test are shown in Fig. \ref{f: amprob}.
The seed for the {\tt Genec} model is built from a polytrope, therefore it does not include the effects of accretion or rotation.
Because of this, when we switch on accretion and rotation at the beginning of the calculation, we let the model redistribute instantly the accreted angular momentum as a solid body for a short time, while thermal adjustment is achieved.
After this, the angular momentum is transported diffusively in the stellar interior and the disk and jet set the rotation rate of the protostar as discussed in Sect. \ref{s: stellar am}.
When the protostar reaches a mass of $2\,\mathrm{M}_\odot$, we switched off the jet and allowed only angular momentum accretion transferred from the Keplerian-like disk.
The result is that the protostellar surface reaches very quickly the critical speed: at the assumed accretion rate of $5\cdot 10^{-4}\,\mathrm{M}_\odot\,\mathrm{yr}^{-1}$, only $\sim 0.1 \,\mathrm{M}_\odot$  could be accreted before the critical limit.
This result supports our conclusion that the angular momentum of the stellar surface is completely dominated by the angular momentum of the accreted material.

The general features of the structure and evolution of the (proto)star are examined in detail in Fig. \ref{f: evolution}.
The Kippenhahn diagram (Fig. \ref{f: evolution}a) reveals the locations and times at which the transfer of energy (and angular momentum) is radiative and convective.
The strong convective motions present in the entirety of the protostar at the beginning of its evolution permit a very efficient redistribution of the angular momentum and produce the solid-body rotation profile seen in Fig. \ref{f: evolution}c for $M_\star = 2.87 \,\mathrm{M}_\odot$.
After $\approx 11.8 \,\mathrm{kyr}$ of evolution, shell deuterium burning begins and with it, the bloating phase.
The star becomes radiative and the angular momentum is not efficiently distributed within the structure, giving rise to the differential rotation profile of Fig. \ref{f: evolution}c for $M_\star = 12\,\mathrm{M_\odot}$.

During the bloating phase, the star spins down (Fig. \ref{f: surf am} lower panel), but because the stellar radius increases, the critical angular velocity also decreases.
After the Kelvin-Helmholtz contraction at $\approx 38\,\mathrm{kyr}$, the star starts burning hydrogen and develops a convective interior.
During this time, the star spins up (although not substantially), but the stellar radius decreases, which means that the critical angular velocity also increases.
A more massive star has a higher critical angular velocity than a lower mass star, so for a fixed criticality, the rotation rate of the star when it reaches hydrogen burning is larger than in the pre-main sequence phase.
At this stage ($M_\star = 40.3\,\mathrm{M_\odot}$ in Fig. \ref{f: evolution}c), the angular momentum structure inside of the star is also distributed depending on whether the regions are convective (leading to flat regions in the curve, that is, solid body rotation) or radiative (where differential rotation can develop).
Our models reveal that despite the changes in radius and mass of the star, a sub-critical rotational equilibrium is found and maintained, and that the rotation criticality is controlled by $\xi$ and $\Lambda$.
This is a consequence of the surface angular momentum being dominated by the angular momentum of the accreting material, and that any spin-up of the star makes $\dot J_\mathrm{jet}$ also more efficient.

\end{appendix}

\end{document}